\documentclass{article}

\usepackage{amsfonts}
\usepackage{epsfig}
\usepackage{amssymb}
\usepackage{amsthm}
\usepackage{newlfont}
\usepackage{amstext}
\usepackage{amsmath}
\numberwithin{equation}{section}

\newtheorem{theorem}{Theorem}
\newtheorem{proposition}{Proposition}
\newtheorem{lemma}{Lemma}

\theoremstyle{definition}

\newtheorem*{remark}{Remark}

\begin{document}

%

%



\title{On $\tau$-function of conjugate nets}

\label{firstpage}

\author{Adam DOLIWA\\ \\ {\it Uniwersytet Warmi\'{n}sko-Mazurski w Olsztynie,
Wydzia{\l} Matematyki i Informatyki}\\
{\it ul.~\.{Z}o{\l}nierska 14 A, 10-561 Olsztyn, Poland}\\ \\
~~E-mail: {\tt doliwa@matman.uwm.edu.pl}}


\date{}

\maketitle

\begin{abstract}
\noindent
We study a potential introduced by Darboux to describe conjugate nets, 
which within the modern theory of
integrable systems can be interpreted as a $\tau$-function.  
We investigate the potential using the non-local $\bar\partial$ dressing method of
Manakov and Zakharov, and we show that it can be interpreted 
as the Fredholm determinant of an integral equation which naturally
appears within that approach. Finally, we give some arguments extending that
interpretation to
multicomponent Kadomtsev--Petviashvili hierarchy.
\end{abstract}

\begin{center}
\hspace{3cm} {\it Dedicated to F. Calogero for his 70-th birthday.}
\end{center}
%
\section{Introduction}

Conjugate nets are certain parametrized submanifolds 
whose theory was the object of investigations of the XIX-th century 
differential geometry \cite{DarbouxOS,Bianchi,Eisenhart-TS}. The basic system of
equations of the theory
\begin{equation} \label{eq:Darboux-h}
\partial_i\partial_j h_k(u) = (\partial_j\log h_i(u)) \partial_i h_k(u) +
(\partial_i\log h_j(u)) \partial_j h_k(u), \quad  
i,j, k \quad \text{distinct},
\end{equation}
constitutes one-half of the system of the Lam\'e equations \cite{Lame}
describing the orthogonal coordinate systems. Here the
functions $h_i$, $i=1,\dots ,N$, $N>2$, called the Lam\'e coefficients, 
depend on parameters $u=(u_1,\dots,u_N)$ of conjugate nets, and
$\partial_i=\frac{\partial}{\partial u_i}$ denote partial derivatives. 
The Darboux system \eqref{eq:Darboux-h} written in terms of the rotation
coefficients $\beta_{ij}$, defined by the compatible system
\begin{equation} \label{eq:adj-lin-h}
\partial_i h_j(u) = h_i(u) \beta_{ij}(u) ,  \quad i\ne j,
\end{equation}
takes the form
\begin{equation} \label{eq:dj-bik}
\partial_j \beta_{ik}(u) = \beta_{ij}(u) \beta_{jk}(u) , \quad j \ne i,k .
\end{equation}
This system is known in the present-day terminology as the $N$-wave 
equation \cite{KvLJMP}, which is the simplest equation of the 
$N$-component ($N>2$)
Kadomtsev--Petviashvili (KP) hierarchy. The parametres $u_i$ of the nets 
can be
identified with the first times of the hierarchy, while the other 
times describe integrable iso-conjugate deformations of the nets \cite{DMMMS}.    

Equations \eqref{eq:dj-bik} imply that one can 
introduce the potentials $\beta_{ii}$, $i=1,\dots,N$, such that
\begin{equation} \label{eq:dj-bii}
\partial_j \beta_{ii}(u) = \beta_{ij}(u) \beta_{ji}(u), \qquad i\ne j .
\end{equation}
The symmetry $i\leftrightarrow j$ of the system \eqref{eq:dj-bii}
allows, in turn, for existence yet another potential
field (this fact was known already to
Darboux, see \cite{DarbouxOS} p. 363)
\begin{equation} \label{eq:b-ii}
\partial_i \log \tau(u) = - \beta_{ii}(u) ,
\end{equation}
which was identified  \cite{DMMMS} with the $\tau$--function of the 
multicomponent
KP hierarchy.  

The $\tau$-functions play the central role
\cite{Sato,DKJM,SegalWilson,GrinevichOrlov,Moerbeke} in
establishing the connections between integrable systems and quantum field
theory, statistical mechanics or the theory of random matrices. They
are often represented in a determinant form or can be identified 
with the Fredholm determinant of the integral Gel'fand--Levitan--Marchenko
equation used to solve the model under consideration. In particular, 
the determinant formula for $\tau$-function of
the KP hierarchy \cite{DKJM} follows from the
free fermions (or ${\mathfrak g}{\mathfrak l}_\infty$) realization of 
the hierarchy. Within the context of the  Zakharov and Shabat dressing
method \cite{ZakharovShabat} the $\tau$-function of the KP hierarchy was
interpreted as the Fredholm determinant in \cite{PoppeSattinger}.

Manakov and Zakharov introduced in \cite{ZakMa} the $\bar\partial$-dressing 
method, based on the nonlocal $\bar\partial$ problem (see also \cite{AYF}) 
as a generalized version of the inverse scattering method \cite{ZMNP}. 
They rediscovered the Darboux system (\ref{eq:Darboux-h}), in the
generalized matrix form, as the "basic set of equations" solvable by the 
$\bar\partial$-dressing method. For example, the 
KP equation \cite{DKJM} was shown in 
\cite{BogdanovManakov} to be a limiting case of the Darboux system.

The multicomponent KP hierarchy and conjugate nets were studied within the 
$\bar\partial$-dressing method in a number of papers 
\cite{BK-mKP,TQL,MMAM-cn}. However, in that approach the determinant 
interpretation of the $\tau$-function of conjugate nets was somehow missing. 
In this paper we show that indeed the
$\tau$-function of conjugate nets can be identified 
with the Fredholm determinant of the integral equation inverting the 
nonlocal $\bar\partial$-problem.

The paper is constructed as follows. In
Section~\ref{sec:dbar-conjnets} we collect the basic elements of the 
$\bar\partial$-dressing method 
and we present the way of solving the Darboux equations within this method.   
Section~\ref{sec:Fdet-tau} is devoted to presentation of the
Fredholm determinant interpretation of the
$\tau$-function of conjugate nets. Finally, in Section \ref{sec:Fdet-tau-mKP}
we briefly discuss generalization of
the above result to the full multicomponent KP hierarchy. 
In the appendix we recall some standard facts \cite{Smithies} 
concerning the Fredholm integral equations.

\section{The $\bar\partial$-dressing method and the Darboux equations}
\label{sec:dbar-conjnets}
In this Section we recall \cite{ZakMa}, \cite{BogdanovManakov}
the method of solution of the Darboux equations
\eqref{eq:Darboux-h} or \eqref{eq:dj-bik} within the $\bar\partial$-dressing 
approach.

The basis of the $\bar\partial$-dressing method is the following
integro-differential equation in the complex plane ${\mathbb C}$
\begin{equation} \label{eq:db-nonl}
\bar\partial (\chi(\lambda) - \eta(\lambda)) + 
(\hat{S}\chi)(\lambda) = 0.
\end{equation}
Here $\hat{S}$ is an integral operator
\begin{equation*}
(\hat{S}\phi)(\lambda) = \int_{\mathbb C} 
S(\lambda,\lambda')\phi(\lambda') \: 
d\lambda'\wedge d\bar\lambda',
\end{equation*}
and the given rational function $\eta(\lambda)$ is called the 
normalization; it is
assumed that 
\begin{equation*}
\chi(\lambda) - \eta(\lambda) \to 0, \qquad 
\text{for} \quad |\lambda|\to\infty.
\end{equation*}

Due to the generalized Cauchy formula 
the solution of the non-local $\bar\partial$ problem~\eqref{eq:db-nonl} 
can be expressed
in terms of the solution of the equation
\begin{equation} \label{eq:int-db}
\chi(\lambda) + 
\frac{1}{2\pi i}\int_{\mathbb C} 
\frac{ (\hat{S}\chi)(\lambda')}{\lambda' - \lambda}
\: d\lambda'\wedge d\bar\lambda' = \eta(\lambda) ,
\end{equation}
which can be put in the form of the Fredholm integral equation of the 
second kind  
\begin{equation} \label{eq:Fredh-db}
\chi(\lambda) = \eta(\lambda) - \int_{\mathbb C} K(\lambda, \lambda'') 
\chi(\lambda'') d\lambda''\wedge d\bar\lambda'',
\end{equation}
with the kernel
\begin{equation} \label{eq:Fredh-ker-db}
K(\lambda, \lambda'') = \frac{1}{2\pi i} \int_{\mathbb C} 
\frac{S(\lambda', \lambda'') }{\lambda' - \lambda}
d\lambda'\wedge d\bar\lambda'  .  
\end{equation}

\begin{remark}
In the paper we always assume that the kernel $S$ in the non-local 
$\bar\partial$
problem is such that the Fredholm equation \eqref{eq:int-db}
is uniquely solvable. Then, by the Fredholm alternative, the homogenous 
equation with $\eta=0$ has only the trivial solution.
\end{remark}

Let us assume that the kernel $S$ in the non-local $\bar\partial$ problem
depends on additional parameters. To get the $\bar\partial$-dressing method 
of construction of solutions fo the Darboux equations 
let us introduce the following simple
dependence of the kernel $S$ on the variables $u_i$, $i=1,\dots,N$
\begin{equation} \label{eq:evol-S-cont}
S(\lambda',\lambda'', u) = g(\lambda',u)^{-1} 
S_0(\lambda',\lambda'') g(\lambda'',u),
\end{equation}
where 
\begin{equation} \label{eq:g-nets}
g(\lambda',u)= \exp \sum_{i=1}^N u_i A_i(\lambda'), 
\qquad A_i(\lambda') =
\frac{c_i}{\lambda' - \lambda_i},
\end{equation}
$c_i$ are non-zero constants, and $\lambda_i\in{\mathbb C}$ 
are points of the
complex plane. Moreover we assume that $\lambda_i\ne\lambda_j$ for
different $i$ and $j$. 

Directly one can verify the following result.
\begin{lemma}
The evolution \eqref{eq:evol-S-cont} of the kernel $S$ implies that the
kernel $K$ of the integral equation \eqref{eq:Fredh-db}
is subjected to the equation 
\begin{equation} \label{eq:evol-K-cont} 
\partial_i K(\lambda,\lambda'',u)  = 
A_i(\lambda) K(\lambda_i,\lambda'',u) +   A_i(\lambda'')
K(\lambda,\lambda'',u) -
A_i(\lambda) K(\lambda,\lambda'',u). 
\end{equation}
\end{lemma}
In consequence of the above formula \eqref{eq:evol-K-cont}
we obtain the following useful and crucial for the $\bar\partial$-dressing 
method result.
\begin{lemma} 
When $\chi(\lambda,u)$ is the unique, by assumption, solution of 
the $\bar\partial$ problem \eqref{eq:db-nonl} with the kernel $S$
evolving according to \eqref{eq:evol-S-cont}, and  
with normalization given by $\eta(\lambda,u)$, then the function
\begin{equation*}
\partial_i \chi (\lambda,u) + A_i(\lambda)\chi (\lambda,u)
\end{equation*}
is the solution of the same $\bar\partial$ problem
but with  the new normalization
\begin{equation*} 
\partial_i\eta(\lambda,u) + A_i(\lambda)\eta (\lambda,u) +
A_i(\lambda)(\chi(\lambda_i,u) - \eta(\lambda_i,u)) .
\end{equation*}
\end{lemma}
The above Lemma leads to the following Theorem, which
gives the system of linear problems for the Darboux equation.
\begin{theorem} \label{th:psi}
Given solution $\chi(\lambda,u)$ of the the $\bar\partial$ problem
\eqref{eq:db-nonl}, \eqref{eq:evol-S-cont} with the normalization
$\eta(\lambda,u)= 1$, then the function
\begin{equation*}
\psi(\lambda,u) = \chi(\lambda,u) g(\lambda,u)
\end{equation*}
satisfies with respect to variables $u_i$
the following system of Laplace equations 
\begin{equation} \label{eq:Lapl-db}
\partial_i\partial_j \psi(\lambda,u) = a_{ij}(u) \partial_i \psi(\lambda,u) + 
a_{ji}(u) \partial_j \psi(\lambda,u) , \quad i\ne j,
\end{equation}
with the coefficients
\begin{equation} \label{eq:aij-db}
a_{ij}(u)  = \frac{\partial_j \chi (\lambda_i,u)}
{\chi(\lambda_i,u)}
+A_j(\lambda_i).
\end{equation}
\end{theorem}
\begin{proof}
The idea of the proof is standard within the $\bar\partial$-dressing method
approach. One collects solutions of the $\bar\partial$ problem 
\eqref{eq:db-nonl} (or, equivalently, 
the integral equation \eqref{eq:Fredh-db}) to obtain
new solution with the vanishing normalization. Then, by the Fredholm
alternative, such solution must be identically zero.
\end{proof}

Define the Lam\'e coefficients $h_i(u)$, by
\begin{equation*}
h_i(u) = \chi(\lambda_i,u) g_i(\lambda_i,u),
\end{equation*}
where
\begin{equation*}
g_i(\lambda,u)=\exp \sum_{j=1,j\ne i}^{N} u_j A_j(\lambda);
\end{equation*}
equivalently, they are the "non-singular" parts of the function 
$\psi(\lambda,u)$ in the points $\lambda_i$
\begin{equation} \label{eq:h-db}
h_i(u) = \lim_{\lambda\to \lambda_i} \left[ \psi(\lambda,u) \; 
\exp (- u_i A_i(\lambda))\right].
\end{equation}
Then the coefficients $a_{ij}(u)$ of the Laplace equations 
\eqref{eq:Lapl-db} read
\begin{equation*}
a_{ij}(u) = \partial_j \log h_i(u) , \qquad i\ne j,
\end{equation*}
and the compatibility condition of \eqref{eq:Lapl-db} is the
Darboux system \eqref{eq:Darboux-h}. Alternatively, the Darboux system
can be obtained evaluating (removing first the singularity like in 
equation \eqref{eq:h-db}) equation 
\eqref{eq:Lapl-db} in the points $\lambda_k$, $k\ne i,j$. 

\begin{remark}
To obtain real solutions one needs special symmetry
properties of the kernel $S$ with respect to the complex conjugation
\begin{equation*}
S(\bar \lambda, \lambda, \bar\lambda',\lambda') = 
\overline{S(\lambda,\bar \lambda,\lambda',\bar\lambda')}, 
\end{equation*}
and reality of the points $\lambda_i=\bar\lambda_i\in{\mathbb R}$, 
and the parameters
$c_i\in{\mathbb R}$, $i=1,\dots,N$,
which imply that
\begin{equation*}
\psi(\bar \lambda,\lambda,u) = \overline{\psi(\lambda,\bar \lambda,u)}, \qquad 
\text{and} \quad h_{i}(u)\in{\mathbb R}.
\end{equation*}
\end{remark}

The following result allows to give the $\bar\partial$-method of 
construction of solutions of the Darboux system \eqref{eq:dj-bik}
written in terms of
the rotation coefficients. Its proof is analogous to the proof of
Theorem \ref{th:psi}.
\begin{theorem}
Let $\chi_i(\lambda,u)$, $i=1,\dots,N$ of the be solutions of
the $\bar\partial$ problem \eqref{eq:db-nonl} with the normalizations
\begin{equation} \label{eq:nu-chi-i}
\eta_i(\lambda,u) = A_i(\lambda) g_i(\lambda_i,u)^{-1},
\end{equation}
then the functions
\begin{equation*}
\psi_i(\lambda,u) = \chi_i(\lambda,u) g(\lambda,u)
\end{equation*}
satisfy equations
\begin{equation} \label{eq:dipsi-db}
\partial_i\psi(\lambda,u) = h_i(u) \psi_i(\lambda,u) .
\end{equation}
Moreover, the functions $\psi_i(\lambda,u)$ satisfy the linear system
\begin{equation} \label{eq:djpsii-db}
\partial_j\psi_i(\lambda,u) = \beta_{ij}(u) \psi_j(\lambda,u) , \quad j\ne i
\end{equation} 
with the rotation coefficients
\begin{equation} \label{eq:bij-db}
\beta_{ij}(u) = \chi_i(\lambda_j,u) g_j(\lambda_j,u)=
\lim_{\lambda\to \lambda_j} \left[ \psi_i(\lambda,u) \; 
\exp (- u_j A_j(\lambda))\right].
\end{equation}
\end{theorem}

Evaluating equations \eqref{eq:dipsi-db} in the points $\lambda_j$, $j\ne i$, 
(removing first the singularity) we obtain equations
\eqref{eq:adj-lin-h}. The compatibility of the linear system 
\eqref{eq:djpsii-db} gives the Darboux equations  \eqref{eq:dj-bik}
in terms of the rotation coefficients. 
Alternatively, equations \eqref{eq:dj-bik}
can be obtained evaluating equations \eqref{eq:bij-db}
in the points $\lambda_k$, $k\ne i,j$.

\section{The first potentials and the $\tau$--function}
\label{sec:Fdet-tau}
To give the meaning of the $\tau$-function within the $\bar\partial$-dressing
method we first present the meaning of the potentials $\beta_{ii}$ defined
by equations \eqref{eq:dj-bii}. 
\begin{proposition}
Within the $\bar\partial$-dressing method the potential $\beta_{ii}(u)$ can be
identified with the non-singular part of the function $\psi_i$ at the point
$\lambda_i$
\begin{equation} \label{eq:bii-lim}
\beta_{ii}(u) = \lim_{\lambda\to \lambda_i} \left( \psi_i(\lambda,u)
\exp(-u_i A_i(\lambda)) - A_i(\lambda)  \right) .
\end{equation}
\end{proposition}
\begin{proof}
Multiplication of both sides of equation \eqref{eq:djpsii-db} by
$g_i(\lambda,u)^{-1}$ and evaluation of the result 
in the limit $\lambda\to\lambda_i$ gives 
\begin{equation*}
\lim_{\lambda\to\lambda_i} \partial_j\left(  \psi_i(\lambda,u)
\exp(-u_i A_i(\lambda)) \right) = \beta_{ij}(u)\beta_{ji}(u).
\end{equation*} 
The expression in brackets in the LHS of the above equation is singular
at $\lambda_i$. Up to a constant term, which vanishes during differentiation,
it agrees with
\begin{equation*}
\psi_i(\lambda,u) \exp(-u_i A_i(\lambda)) - A_i(\lambda) ,
\end{equation*}
which is finite at $\lambda_i$, due to the normalization condition 
\eqref{eq:nu-chi-i}.
After this replacement we can exchange differentiation with taking the
limit.
\end{proof}

Finally, we give the meaning to the next potential, the $\tau$-function,
which is related with potentials
$\beta_{ii}(u)$ by equations \eqref{eq:b-ii}.  
\begin{theorem} \label{th:tau=Fdet}
Within the $\bar\partial$-dressing method, the $\tau$-function of conjugate nets
can be identified with the Fredholm determinant of the integral equation
\eqref{eq:Fredh-db} inverting the non-local $\bar\partial$-problem 
\eqref{eq:db-nonl} with kernel evolving according to equation
\eqref{eq:evol-S-cont}.
\end{theorem}
Before proving this theorem we first show the following Lemma; we
use here the notation of Appendix \ref{app:Fredholm}.
\begin{lemma} \label{lem:di-KK}
\begin{equation*}
\partial_i K \left( \begin{matrix} z_1 &  \dots  & z_m \\ 
z_1 & \dots  & 
z_m \end{matrix}\; \biggr\rvert \; u \right) = \sum_{\ell = 1}^{m} A_i(z_\ell)
K \left( \begin{matrix} 
\lambda_i & z_1 &  \dots & \check{z}_\ell & \dots & z_m \\
z_\ell & z_1 & \dots  & \check{z}_\ell & \dots & z_m \end{matrix}
\; \biggr\rvert \; u \right),
\end{equation*}
where the symbol $\check{z}_\ell$ means that $z_\ell$ should be removed
from the sequence.
\end{lemma}
\begin{proof}
Denote by $\pi_m$ the set of all
permutations of $\{ 1, \dots, m \}$.
Differentiation of the expression  
\begin{equation*}
K \left( \begin{matrix} z_1 & \dots  & z_m \\ z_1 & \dots  & 
z_m \end{matrix} \; \biggr\rvert \; u \right) = 
\sum_{\sigma \in \pi_m} \text{sgn} \sigma \;
K(z_1,z_{\sigma(1)},u) \cdot \hdots \cdot
K(z_m,z_{\sigma(m)},u) \; ,
\end{equation*}
with respect to $u_i$ gives, by application of equation 
\eqref{eq:evol-K-cont}, 
\begin{equation*}
\partial_i K \left( \begin{matrix} z_1 & \dots  & z_m \\ 
z_1 & \dots  & 
z_m \end{matrix} \; \biggr\rvert \; u \right) = 
\sum_{\ell = 1}^m \sum_{\sigma \in \pi_m} \text{sgn} 
\sigma \;
K(z_1,z_{\sigma(1)},u)   \cdot \hdots 
\cdot A_i(z_\ell)K(\lambda_i,z_{\sigma(\ell)},u)  \cdot \hdots \cdot 
K(z_m,z_{\sigma(m)},u) \; ,
\end{equation*}
where we have used also the following 
elementary formula valid for any permutation 
$\sigma\in\pi_m$
\begin{equation*}
\sum_{\ell = 1}^m A_i(z_\ell) = \sum_{\ell = 1}^m A_i(z_{\sigma(\ell)}) .
\end{equation*}
After application of an even number of transpositions in any of the $m$ 
determinants we obtain the statement of the Lemma.
\end{proof}
\begin{proof}[Proof of Theorem \ref{th:tau=Fdet}]
Using Lemma \ref{lem:di-KK} we can derive the following formula for
$i$-th derivative of the Fredholm determinant $D_F(u)$ 
\begin{equation*}
\partial_i D_F (u) = \sum_{m=1}^{\infty} \frac{1}{(m-1)!}\int_{{\mathbb C}^{m}}
A_i(\lambda') K \left(\begin{matrix} 
\lambda_i & z_1 &  \dots  & z_{m-1} \\
\lambda' & z_1 & \dots   & z_{m-1} \end{matrix}\; \biggr\rvert \; u \right) 
d\lambda'\wedge d\bar\lambda'\dots
d{z_{m-1}}\wedge d\bar{z}_{m-1} \; . 
\end{equation*}
Comparing with
\begin{equation*}
D_F(\lambda_i,\lambda',u) = \sum_{m=0}^{\infty} \frac{1}{m!}\int_{{\mathbb C}^{m}} 
K \left(\begin{matrix} 
\lambda_i & z_1 &  \dots  & z_m \\
\lambda' & z_1 & \dots   & z_m \end{matrix}\; \biggr\rvert \; u\right) 
d z_1\wedge d\bar{z}_1 \dots d{z_m}\wedge d\bar{z}_m \;
\end{equation*}
we obtain that 
\begin{equation*}
\partial_i D_F(u) = \int_{\mathbb C} D_F(\lambda_i,\lambda',u) A_i(\lambda') 
d\lambda'\wedge d\bar\lambda' .
\end{equation*}
Notice that the solution of the Fredholm equation \eqref{eq:Fredh-db}
with normalization $A_i(\lambda)$ is given, due to \eqref{eq:nu-chi-i} and  
\eqref{eq:Fr-sol}, by
\begin{equation*}
\chi_i(\lambda,u) g_i(\lambda_i,u) = A_i(\lambda) - \int_{\mathbb C} 
\frac{D_F(\lambda,\lambda',u)}{D_F(u)} A_i(\lambda') 
d\lambda'\wedge d\bar\lambda' \; .
\end{equation*}
In the limit $\lambda\to\lambda_i$ (compare with \eqref{eq:bii-lim})
we obtain, therefore,
\begin{equation*}
\partial_i \log D_F(u) = - \beta_{ii}(u) \; ,
\end{equation*}
which allows, via equation \eqref{eq:b-ii}, for identification of the 
$\tau$-function with the Fredholm determinant $D_F(u)$.
\end{proof}

\section{Conclusion and remarks}
\label{sec:Fdet-tau-mKP}
We have shown that within the $\bar\partial$-dresing method
the $\tau$-function of conjugate nets can be identified with the Fredholm
determinant inverting the non-local $\bar\partial$ problem. In fact, 
this result can be extended to the full $N$-component KP hierarchy. 
To justify this opinion we should: 
\begin{enumerate}
\item incorporate higher times of the hierarchy into the scheme,
\item give the meaning to the full set of $\tau$-functions labelled by the 
root lattice vectors of the $A_{N-1}$ root system. 
\end{enumerate}
The first task can be done by using the idea of 
\cite{BogdanovManakov}, where the KP equation was obtained from the Darboux 
system (the scalar "basic set of equations") for $N=3$. In the same way 
the higher times can be included into evolution of the
kernel. The second problem
can be solved by combination of the result of \cite{DMMMS} on relation of the
Schlesinger transformations on the $A_{N-1}$ root lattice with the 
Laplace transformations of conjugate nets, with the construction of such 
Laplace transformations within the $\bar\partial$-dresing method given in
\cite{TQL}. The details will be presented elsewhere.

\appendix

\section{Elements of the Fredholm theory}
\label{app:Fredholm}
We recall in this Appendix some standard facts (see, for example
\cite{Smithies}) from the theory of Fredholm
integral equations which we use in the paper. 

Consider the Fredholm equation of the second kind 
\begin{equation}  \label{eq:Fr-eq}
f(x) = g(x) - \int_\Omega K(x,y) f(y) dy \; ,
\end{equation}
where $K(x,y)$ is the kernel of the integral
operator, and $g(x)$ is a given function. 
The Fredholm determinant $D_F$ is defined by the series
\begin{equation}  \label{eq:Fr-det}
D_F = 1 +
\sum_{m=1}^\infty \frac{1}{m!} \int_{\Omega^m}
K \begin{pmatrix} x_1 & x_2 & \dots  & x_m \\ x_1 & x_2 & \dots  & x_m
\end{pmatrix} dx_1 \dots dx_m \; ,
\end{equation}
where
\begin{equation*}  \label{eq:Fr-KK}
K \begin{pmatrix} x_1 & x_2 & \dots  & x_m \\ y_1 & y_2 & \dots  & y_m
\end{pmatrix} = \det \begin{pmatrix} 
K(x_1,y_1) & K(x_1,y_2) & \dots  & K(x_1,y_m)  \\ 
\vdots & \vdots & & \vdots \\
K(x_m,y_1) & K(x_m,y_2) & \dots  & K(x_m,y_m)
\end{pmatrix} ,
\end{equation*}
while the Fredholm minor is defined by the series
\begin{equation}  \label{eq:Fr-min}
D_F(x,y) = \sum_{m=0}^\infty \frac{1}{m!} 
\int_{\Omega^m} K \begin{pmatrix} 
x & x_1 & x_2 & \dots  & x_m \\ y & x_1 & x_2 & \dots  & x_m
\end{pmatrix} dx_1 \dots dx_m \; .
\end{equation}
For non-vanishing Fredholm determinant
$D_F$ the solution of the integral equation \eqref{eq:Fr-eq} 
is unique and can be written in 
the form
\begin{equation}  \label{eq:Fr-sol}
f(x) = g(x) - \int_\Omega \frac{D_F(x,y)}{D_F} 
g(y) dy \; .
\end{equation}

\section*{Acknowledgments}
The paper was partially supported by the 
University of Warmia and Mazury in Olsztyn under the grant  522-1307-0201
and by KBN grant 2-P03B-12-622.

\label{lastpage}


\begin{thebibliography}{99}
\small

\bibitem{AYF}
Ablowitz M J, Bar Yaacov D and Fokas A S, On the inverse scattering
problem for the {Kadomtsev}--{Petviashvili} equation, {\it Stud. Appl.
Math.} {\bf 69} (1983), 135--143.

\bibitem{Bianchi}
Bianchi L, Lezioni di geometria differenziale, Zanichelli, Bologna, 1924.

\bibitem{BK-mKP}
Bogdanov L V and Konopelchenko B G, Analytic-bilinear approach to 
integrable hierarchies {II}.  {Multicomponent} {KP} and {2D} {Toda} 
lattice hierarchies, {\it J. Math. Phys.} {\bf 39} (1998), 4701--4728.

\bibitem{BogdanovManakov}
Bogdanov L V and Manakov S V, The nonlocal $\bar{\partial}$-problem
and (2+1)-dimensional soliton equations, {\it J. Phys. A: Math. Gen.} 
{\bf 21} (1988), L537--L544.

\bibitem{DarbouxOS}
Darboux G, Le\c{c}ons sur les syst\'emes orthogonaux et les coordonn\'ees
curvilignes, Gauthier -- Villars, Paris, 1910.

\bibitem{DKJM}
Date E, Kashiwara M, Jimbo M, and Miwa T, Transformation groups for
soliton equations, in Nonlinear integrable systems --- classical theory and
quantum theory, Proc. of RIMS Symposium, Editors: Jimbo M and Miwa T, World
Scientific, Singapore, 1983, 39--119.

\bibitem{TQL}
Doliwa A, Santini P M and Ma{\~n}as M, Transformations of
quadrilateral lattices, {\it J. Math. Phys.} {\bf 41} (2000), 944--990.


\bibitem{DMMMS}
Doliwa A, Ma{\~n}as M, Mart{\'\i}nez Alonso L, Medina E and Santini P M,
Multicomponent {KP} hierarchy and classical transformations of
conjugate nets, {\it J. Phys. A} {\bf 32} (1999), 1197--1216.

\bibitem{Eisenhart-TS}
Eisenhart L P, Transformations of surfaces, Princeton University
Press, Princeton, 1923.

\bibitem{GrinevichOrlov}
Grinevich P G and Orlov A Yu, Flag spaces in {KP} theory and
{V}irasoro action on $\det \bar\partial_j$ and {S}egal--{W}ilson
$\tau$-function, in Problems of Modern Quantum Field Theory, 
Springer-Verlag, Berlin, 1989, 86--106.

\bibitem{KvLJMP}
Kac V G and ~van~de Leur J, The n-component {KP} hierarchy and
representation theory, {\it J. Math. Phys.} {\bf 44} (2003), 3245--3293.


\bibitem{Lame}
Lam\'e G, Le\c{c}ons sur les coordonn\'ees curvilignes et leurs diverses
applications, Mallet--Bachalier, Paris, 1859.

\bibitem{MMAM-cn}
Ma{\~n}as M, Mart{\'\i}nez Alonso L and Medina E, Dressing methods for geometric
nets:~I. Conjugate nets, {\it J. Phys. A: Mathematical and General} {\bf 33}
(2000) 2871--2894.

\bibitem{Moerbeke}
van Moerbeke P, Integrable Lattices: Random matrices and Random Permutations, in
Random Matrices and their Applications, MSRI publications {\bf 40} (2001),
321-406 ({\tt math.CO/0010135}). 

\bibitem{PoppeSattinger}
P\"{o}ppe Ch and Sattinger D H, Fredholm determinants and the $\tau$
function for the Kadomtsev--Petviashvili hierarchy, {\it Publ. RIMS, Kyoto
Univ.} {\bf 24} (1988), 505--538.


\bibitem{Sato}
Sato M, Soliton equations as dynamical systems on infinite dimensional
{G}rassman manifolds, {\it RIMS Kokyuroku} {\bf 439} (1981), 30--46.

\bibitem{SegalWilson}
Segal G and Wilson G, Loop groups and equations of {K}d{V} type, {\it Inst.
Hautes \'Etudes Sci. Publ. Math.} {\bf 61} (1985), 5--65.

\bibitem{Smithies}
Smithies F, Integral Equations, Cambridge Univ. Press, Cambridge, 1965.

\bibitem{ZakMa}
Zakharov V E and Manakov S V, Construction of multidimensional
nonlinear integrable systems and their solutions, {\it Funct. Anal. Appl.}
{\bf 19} (1985), 11--25.

\bibitem{ZakharovShabat}
Zakharov V E and Shabat A B, A scheme for integrating the nonlinar equations
of mathematical physics by the method of the inverse scattering problem, 
{\it Funct. Anal. Appl.} {\bf 8} (1974), 226-235.

\bibitem{ZMNP}
Zakharov V E, Manakov S V, Novikov S P and Pitaevskii L P, Theory of
solitons --- inverse scattering metod, Nauka, Moscow, 1980.




\end{thebibliography}
\end{document}